\documentclass[preprint,aps,prc,superscriptaddress,floatfix,showpacs]{revtex4}

\usepackage{graphicx}
\usepackage{bm}
\usepackage{CJK}
\usepackage{color}
\begin{document}
\newcommand{\ket}[1]{|#1\rangle}
\newcommand{\Phirm}{\mathrm{\Phi}}
\title{Triaxial superdeformation in $^{40}$Ar}
\begin{CJK*}{JIS}{}
\author{Yasutaka~Taniguchi}
\affiliation{RIKEN Nishina Center for Accelerator-Based Science, RIKEN, Wako, Saitama 351-0198, Japan}

\author{Yoshiko~Kanada-En'yo}
\affiliation{Yukawa Institute for Theoretical Physics, Kyoto University, Kyoto, Kyoto 606-8502, Japan}

\author{Masaaki~Kimura}
\affiliation{Creative Research Initiative ``Sousei,'' Hokkaido University, Sapporo, Hokkaido 001-0021, Japan}

\author{Kiyomi~Ikeda}
\affiliation{RIKEN Nishina Center for Accelerator-Based Science, RIKEN, Wako, Saitama 351-0198, Japan}

\author{Hisashi~Horiuchi}
\affiliation{Research Center for Nuclear Physics, Osaka University, Ibaraki, Osaka 567-0047, Japan}

\author{Eiji~Ideguchi}
\affiliation{Center for Nuclear Study, The University of Tokyo, Wako, Saitama 351-0198, Japan}
\date{\today}
\begin{abstract}
Superdeformed (SD) states in $^{40}$Ar have been studied using the deformed-basis antisymmetrized molecular dynamics. 
Low energy states were calculated by the parity and angular momentum projection (AMP) and the generator coordinate method (GCM).
Basis wave functions were obtained by the energy variation with a constraint on the quadrupole deformation parameter $\beta$, while other quantities such as triaxiality $\gamma$ were optimized by the energy variation. 
 By the GCM calculation, an SD band was obtained just above the ground state (GS) band. 
 The SD band involves a $K^\pi = 2^+$ side band due to the triaxiality. 
The calculated electric quadrupole transition strengths of the SD band reproduce the experimental values appropriately.
Triaxiality is significant for understanding low-lying states. 
 \end{abstract}
\maketitle
\end{CJK*}

Numerous superdeformed (SD) bands have been identified and studied in various domains of the nuclear chart. 
An example of recent progress in this research area is the discovery of the SD states in a very light mass region ($A\sim 40$) such as $^{36}$Ar \cite{PhysRevC.63.061301}, $^{40}$Ca \cite{PhysRevLett.87.222501}, and $^{44}$Ti \cite{PhysRevC.61.064314}. 
A striking feature of these experimental works is that the SD band members were assigned from band heads up to high spin states.
Discrete $\gamma$ transitions linked those SD bands to spherical or normal-deformed states, which established the excitation energies and the spin-parities of the SD bands.

Theoretical microscopic studies of deformed states in the $A\sim 40$ region have been performed based on the shell model \cite{PhysRevLett.95.042502,PhysRevC.75.054317}, mean-field
models \cite{Inakura2002261,R.R.Rodriguez-Guzman2004,PhysRevC.68.044321}, and the
antisymmetrized molecular dynamics (AMD)
\cite{Kimura200658,taniguchi:044317,PhysRevC.69.051304,PhysRevC.72.064322,PTP.121.533,PhysRevC.80.044316}.
Calculations with no assumption for the axial
symmetry suggest the systematic triaxial deformation of the SD states. 
For example, the AMD calculations suggest that SD bands in $^{40}$Ca \cite{taniguchi:044317} and $^{44}$Ti \cite{Kimura200658} are constructed from triaxial shapes and their side bands exist.
The triaxiality of superdeformations in this mass region has also been discussed in the three-dimensional coordinate-mesh Hartree-Fock (HF) \cite{Inakura2002261}.

Triaxiality is also an important feature of normal-deformed states in the $A\sim 40$ region.
Established triaxial states are in $^{40}$Ca, in which an observed $K^\pi = 2^+$ band with low excitation energies \cite{PhysRevLett.87.222501} is interpreted as a side band of a $K^\pi = 0^+$ band caused by the triaxial normal-deformation \cite{Gerace1969241,Gerace1977253,taniguchi:044317,PhysRevC.75.054317}. 
Therefore, triaxiality is one of the key properties to clarify the structure of normal-deformations and superdeformations in $A\sim 40$ nuclei. However, theoretical studies taking into account triaxiality are still limited due to the higher numerical costs than axially symmetric calculations.

Quite recently, an SD $K^\pi = 0^+$ band built on the $J^\pi = 0^+$ state (2.12
MeV) in $^{40}$Ar was experimentally identified up to the $J^\pi = (12^+)$ state by
the $^{26}$Mg($^{18}$O,~2p2n)$^{40}$Ar reactions \cite{ideguchi2010}, which were the
lightest SD states ever in $N \neq Z$ nuclei. Superdeformations in $N \neq Z$ nuclei in this mass region are frontier topics.

To discuss triaxiality, features of unnatural spin-parity [$\pi = (-)^{J+1}$] states are helpful data. In $^{40}$Ar, a state at 4.23 MeV was assigned to an unnatural parity $J^\pi = 3^+$ state by the $^{40}$Ar($\alpha$, $\alpha'$) reaction \cite{PhysRevC.20.38}. 
This $J^\pi = 3^+$ state decays to the $J^\pi = 2^+$ state in the SD band \cite{Southon1976263,Bitterwolf1983}. 
The $J^\pi = 2^+$ state at 3.92 MeV just below the $J^\pi = 3^+$ at 4.23 MeV also decays to the SD band member $J^\pi = 0^+$, and the $E2$ transition strength takes a non-negligible value \cite{Bitterwolf1983}. 
These observations show that the states $J^\pi = 2^+$ and $3^+$ are candidates for members of the side band in the SD band, which can be discussed in relation to the triaxiality of the superdeformation.
Nevertheless, neither theoretical studies of the SD state nor the triaxiality of $^{40}$Ar have progressed.

The purpose of this study is to investigate the structure of the SD band in $^{40}$Ar focusing on triaxiality. 
Competitions of deformations favored by protons and neutrons are also discussed. 

To obtain the wave functions in low-lying states, the parity and angular momentum projection (AMP) and the generator coordinate method (GCM) with deformed-basis AMD wave functions were performed. 
A deformed-basis AMD wave function $\ket{\Phirm}$ is a Slater determinant of triaxially deformed Gaussian wave packets, such that 
\begin{eqnarray}
 \ket{\Phirm} &=& \hat{\cal A}\ket{\varphi_1, \varphi_2, \cdots, \varphi_A}, \\
 \ket{\varphi_i} &=& \ket{\phi_i} \otimes \ket{\chi_i} \otimes \ket{\tau_i}, \\
 \langle \mathbf{r} | \phi_i \rangle &=& \prod_{\sigma = x, y, z} \left( \frac{2 \nu_\sigma}{\pi}\right)^\frac{1}{4} \exp\left[ - \nu_\sigma \left( \sigma - \frac{Z_{i\sigma}}{\sqrt{\nu_\sigma}}\right)^2\right], \nonumber\\
 \ \\
 \ket{\chi_i} &=& \chi^\uparrow_i \ket{\uparrow} + \chi^\downarrow_i \ket{\downarrow},\ \ket{\tau_i} = \ket{\pi}\ \mathrm{or}\ \ket{\nu},
\end{eqnarray}
where $\hat{\cal A}$ is the antisymmetrization operator and $\ket{\varphi_i}$ are single-particle wave functions. 
The variables $\ket{\phi_i}$, $\ket{\chi_i}$, and 
$\ket{\tau_i}$ are a spatial part, a spin part, and an isospin part, respectively, of every single-particle wave function $\ket{\varphi_i}$.
The values $\bm{\nu} = (\nu_x, \nu_y, \nu_z)$ are real parameters that indicate the width of the Gaussian single-particle wave function, and $\mathbf{Z}_i = (Z_{ix}, Z_{iy}, Z_{iz})$ are complex parameters that indicate centroids of the single-particle wave functions. The width parameters $\bm{\nu}$ are common to all nucleons. 
The complex parameters $\chi_i^\uparrow$ and $\chi_i^\downarrow$ represent spin directions.
Basis wave functions of the GCM were obtained by the energy variation with a constraint on the quadrupole deformation parameter $\beta$ of a total system after the projection to positive parity states.
Variation parameters were $\bm{\nu}$, $\mathbf{Z}_i$, and $\chi_i^{\uparrow,\downarrow}$.
The isospin part of each single-particle wave function was fixed as a proton or a neutron. 
The Gogny D1S force was used as an effective interaction.
Details of the framework are provided in Refs.~\onlinecite{PhysRevC.56.1844,PhysRevC.69.044319,PTP.93.115}. 

\begin{figure}[tbp]
 \includegraphics[width=0.5\textwidth]{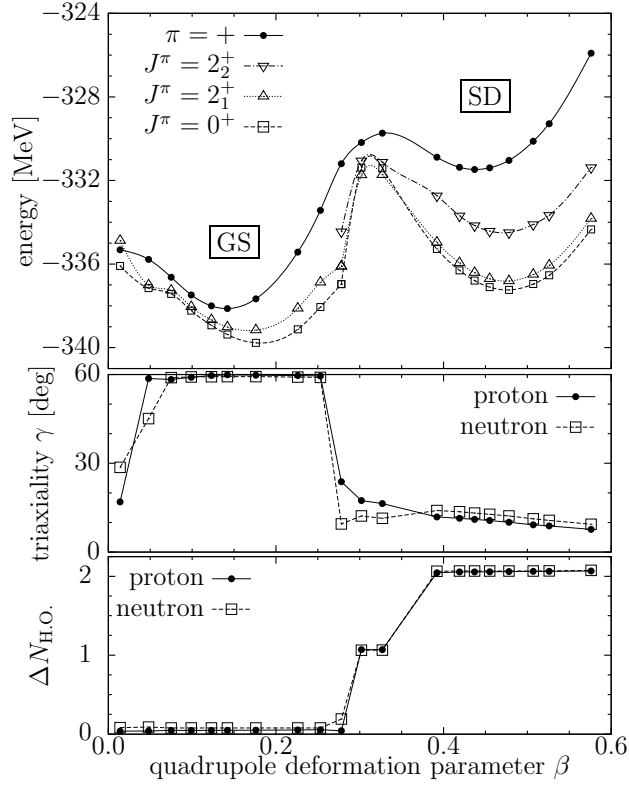}
 \caption{(upper) Energies projected to positive parity states, $J^\pi = 0^+$ and $2^+$ states, (middle) the triaxiality $\gamma$ for protons and neutrons, and (lower) the H.O. values relative to lowest allowed states are plotted as functions of the quadrupole deformation parameter $\beta$ of the total system. }
 \label{fig:beta-energy}
\end{figure}

Calculated values of energy, triaxiality $\gamma$, and harmonic oscillator (H.O.) quanta $\Delta N_\mathrm{H.O.}$ for protons and neutrons measured from the lowest allowed states are shown as functions of the quadrupole deformation parameter $\beta$ in the upper, middle, and lower panels, respectively, of Fig.~\ref{fig:beta-energy}.
As for the energy curves, energies of the positive parity states and $J^\pi = 0^+$ and $2^+$ states, which were projected from each basis wave function, are plotted.
$K$-mixing was performed for the $J^\pi = 2^+$ states.
Those energy curves have two local minima at $\beta \sim 0.2$ and 0.5, which are labeled GS and SD minima, respectively, 
because the GCM results show that these minima correspond to the GS and the SD states, respectively, as explained later. 
In the energy curves projected to positive parity states, the energy gap between the GS and SD minima is approximately 6.7 MeV, while the gap decreases to 2.5 MeV after the AMP to $J^\pi = 0^+$ states. 

\begin{figure}[tbp]
 \begin{tabular}{cc}
  \includegraphics[width=0.225\textwidth]{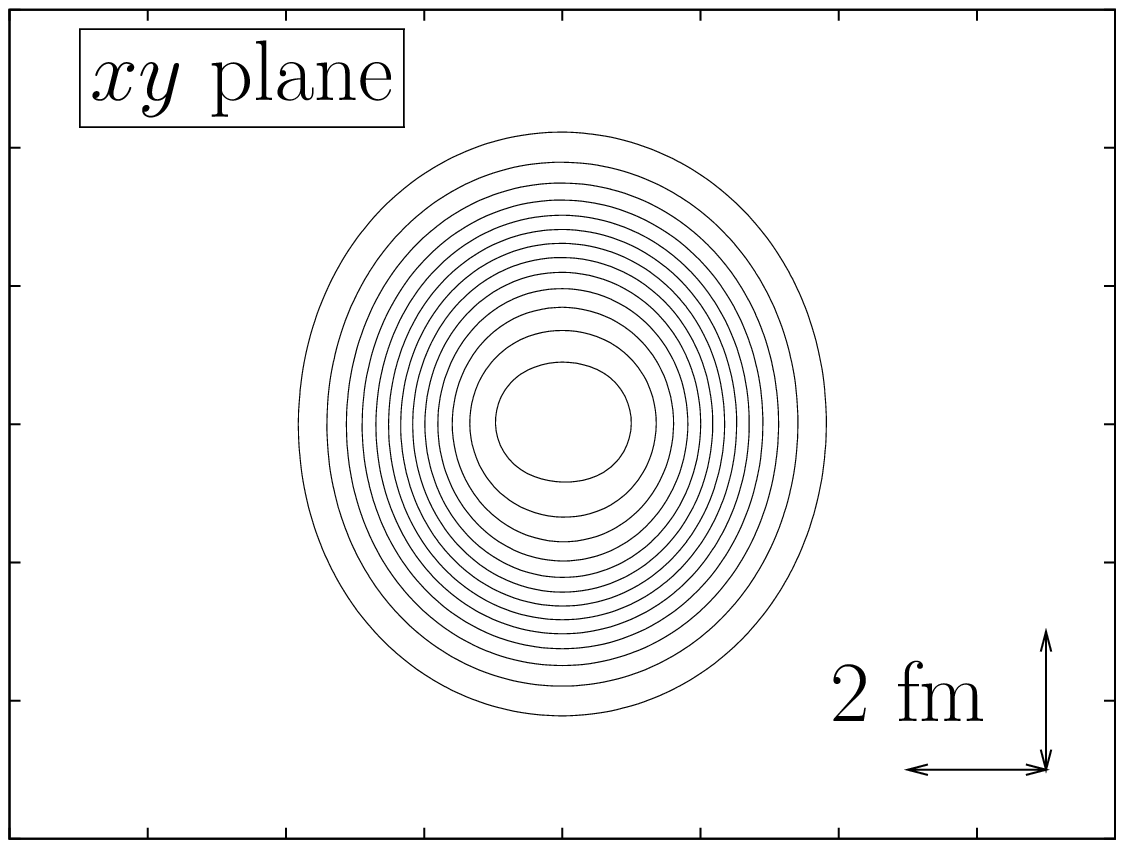} & 
  \includegraphics[width=0.225\textwidth]{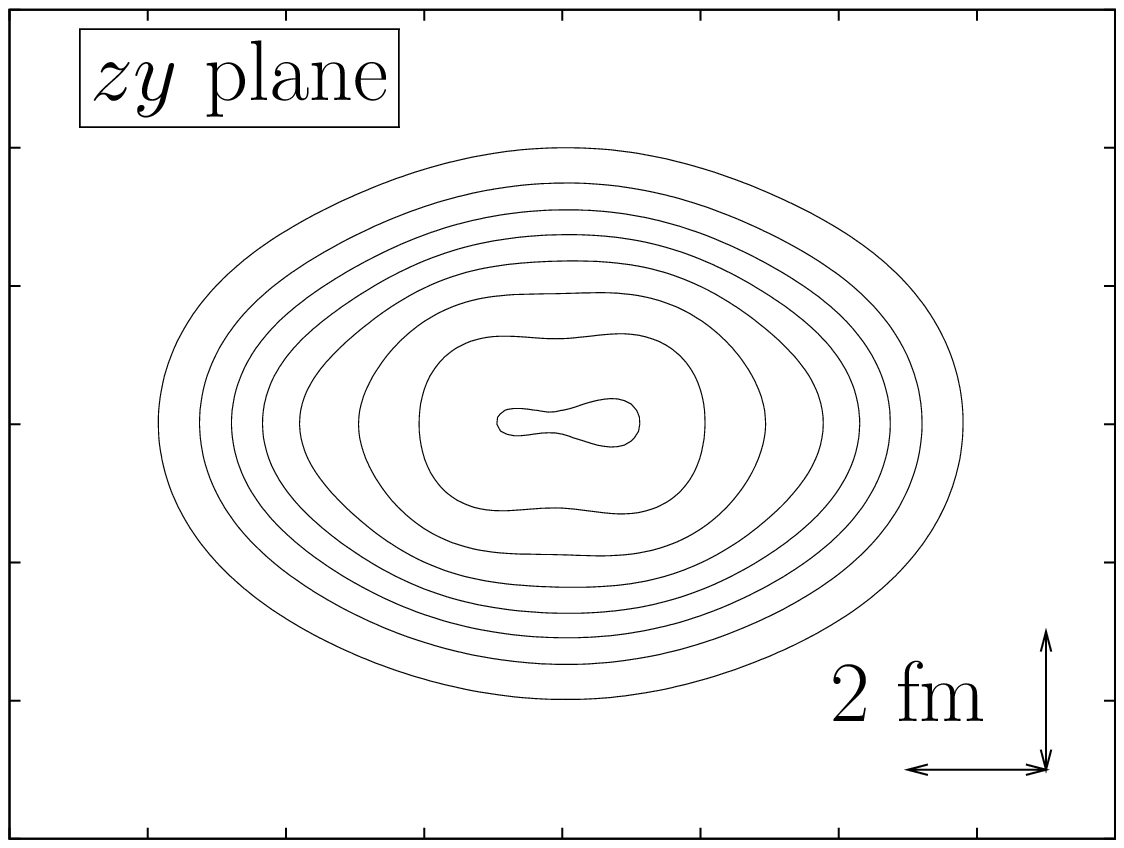} \\
  \includegraphics[width=0.225\textwidth]{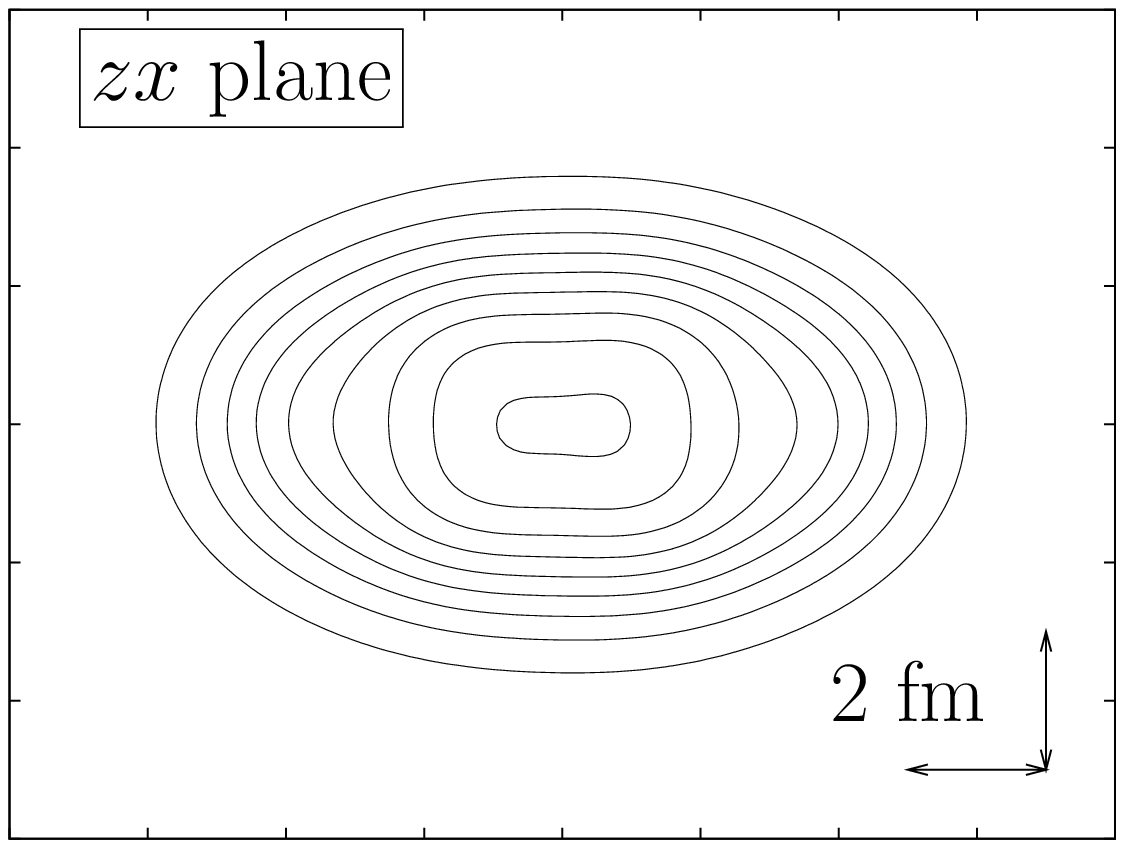}
 \end{tabular}
 \caption{Density distributions of the obtained wave functions with $(\beta, \gamma) = (0.478, 11.2^\circ)$ projected onto the $xy$, $zy$, and $zx$ planes. }
 \label{density}
\end{figure}

The values of triaxiality $\gamma$ for protons and those for neutrons are similar in the entire $\beta$ region (the middle panel of Fig.~\ref{fig:beta-energy}). 
In the small $\beta$ region, the system forms an oblate shape with $\gamma \sim 60^\circ$, while in the large $\beta$ region around the SD minimum, it forms a triaxial deformation with $\gamma \sim 10^\circ$.
Density distributions of the obtained wave function at the SD minimum [$(\beta, \gamma) = (0.478, 11.2^\circ)$] are shown in Fig.~\ref{density}. 
Because of triaxial deformations around the SD minimum, each wave function contains different $K$ components, and two $J^\pi =2^+$ states are obtained after diagonalizing those different $K$ components, i.e., $K$-mixing (the upper panel of Fig.~\ref{fig:beta-energy}).

 The relative H.O. quanta $\Delta N^{\pi}_\mathrm{H.O.}$ and $\Delta N^{\nu}_\mathrm{H.O.}$ for protons and neutrons, respectively, as functions of the quadrupole deformation parameter $\beta$ (the lower panel of Fig.~\ref{fig:beta-energy}) show the structural changes along $\beta$ in terms of particle-hole excitations across major shell gaps. 
 The $\Delta N^{\pi, \nu}_\mathrm{H.O.}$ are defined as
  \begin{eqnarray}
   &&\Delta N^{\pi, \nu}_\mathrm{H.O.}\nonumber\\
   &&= \sum_{\sigma = x, y, z} \left.\left\langle \sum_{i \in \pi, \nu} \frac{\hat{p}_{i\sigma}^2}{2m} + \frac{1}{2} m \omega^2_\sigma \sigma^2 \right\rangle\right/ \hbar \omega_\sigma - \frac{3}{2} A \nonumber\\
   &&\ \ \ \ \ \ \ \ \ \ \ \ \ \ \ \ \ \ \ \ \ \ \ \ \ \ \ \ \ \ \ \ \ \ \ \ \ \ \ \ \ \ \ \ \ \ \ \ - N^{\pi, \nu}_\mathrm{min}.
  \end{eqnarray}
Here $m$ is the nucleon mass, and the frequency ${\bm \omega} = (\omega_x, \omega_y, \omega_z)$ of the H.O. was chosen to be the deformed one ${\bm \omega} = \frac{2 \hbar}{m} {\bm \nu}$ with the width parameters ${\bm \nu}$ of each wave function. 
$N^\pi_\mathrm{min}$ and $N^\nu_\mathrm{min}$ are H.O. quanta of the lowest allowed state for protons and neutrons, respectively, and in the case of $^{40}$Ar, $(N^\pi_\mathrm{min}, N^\nu_\mathrm{min}) = (26, 36)$.
In the small $\beta$ region, the relative H.O. quanta $\Delta N_\mathrm{H.O.}$ take an approximately zero value and indicate no particle-hole excitation for both protons and neutrons.
On the other hand, in the $\beta\gtrsim 0.4$ region, the values of $\Delta N_\mathrm{H.O.}$ jump to two for both protons and neutrons. This shows that the SD minimum has $2 \hbar \omega$ excited configurations for both protons and neutrons, which is a total of $4 \hbar \omega$ excited configurations. 
In $\beta \sim 0.3$, $(\Delta N_{\rm H.O.}^\pi, \Delta N_{\rm H.O.}^\nu) = (1, 1)$ components are obtained.

\begin{figure}[tbp]
  \includegraphics[width=0.5\textwidth]{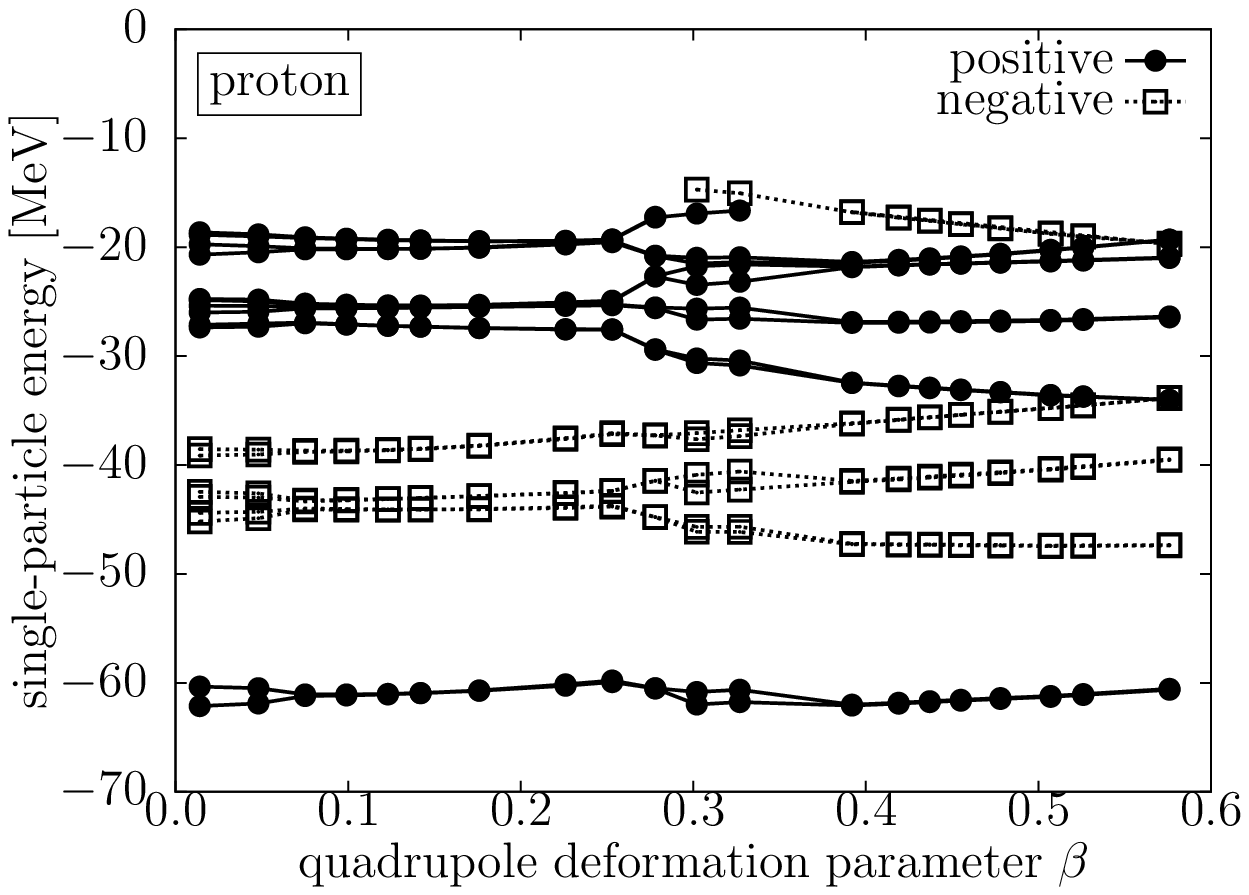}
  \includegraphics[width=0.5\textwidth]{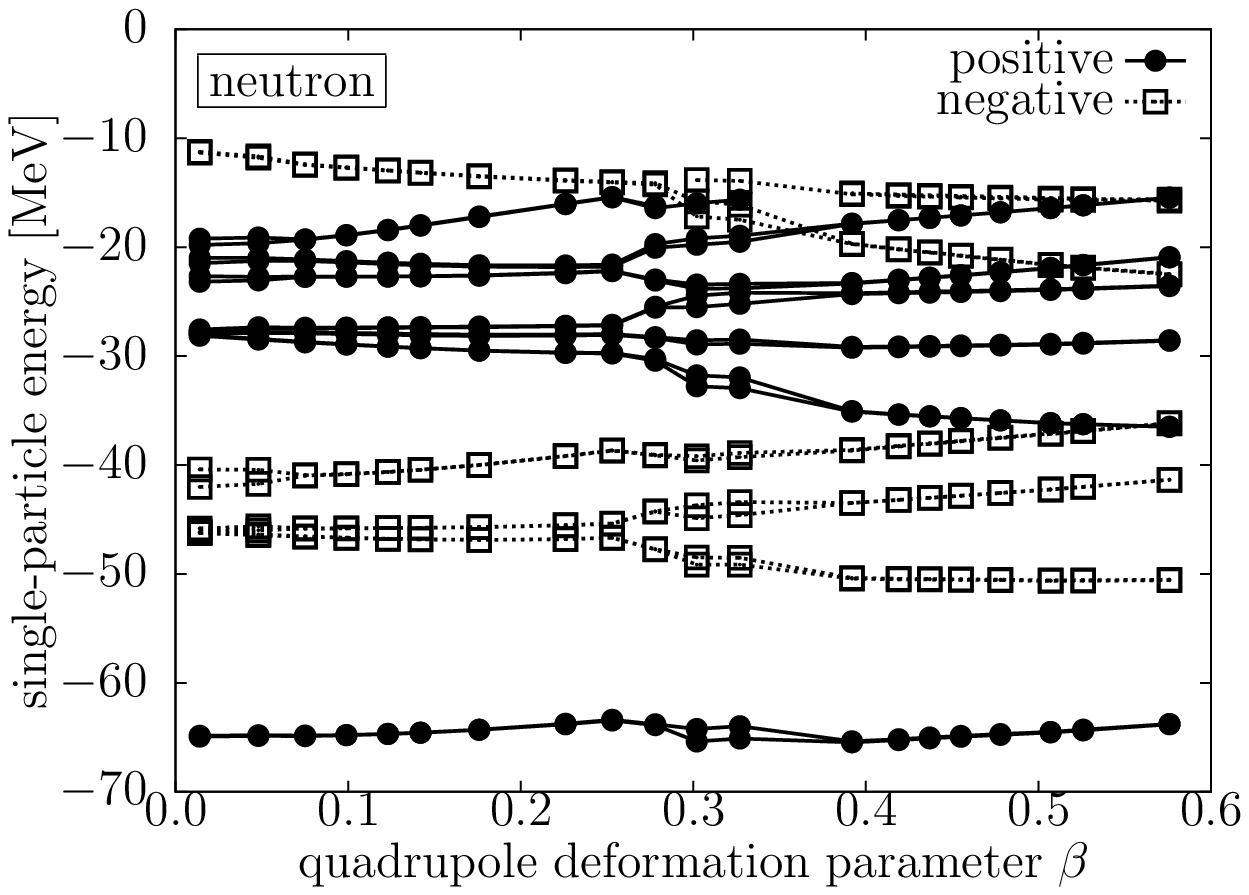}
 \caption{Single-particle energies for protons (upper) and neutrons (lower) are plotted as functions of the quadrupole deformation parameter $\beta$ of the total system.}
 \label{fig:spo}
\end{figure}

To investigate the single-particle behavior in detail, energies and parities of single-particle orbits of the obtained basis wave functions were calculated by transforming the deformed-basis AMD single-particle wave functions to the HF single-particle orbits \cite{PhysRevC.56.1844}. Since most of the obtained wave functions have approximately parity symmetric shapes, extracted single-particle orbits are almost pure-parity orbits.
Figure \ref{fig:spo} shows single-particle energies of positive- and negative-parity orbits for protons and neutrons. 
Two orbits degenerate because of the time reversal symmetry except for $\beta \sim 0$ and 0.3. 
In the small $\beta$ region with $\beta \lesssim 0.25$, the single-particle orbits are normal order for both protons and neutrons, which shows that configurations around the GS minimum are roughly $[(sd)^{-2}]_\pi [(pf)^2]_\nu$ relative to the ground state of $^{40}$Ca. 
With an increase of deformation $\beta$, the highest positive-parity orbits in the $sd$-shell disappear at $\beta \sim 0.3$ and negative-parity orbits come down for both protons and neutrons, which corresponds to the orbit inversion when the $pf$-orbits intrude into the $sd$-orbits in the large $\beta$ region.
Consequently, configurations in the large $\beta$ region with $\beta \gtrsim 0.35$ around the SD minimum are roughly $[(sd)^{-4}(pf)^2]_\pi [(sd)^{-2}(pf)^{4}]_\nu$. 

\begin{figure*}[tbp]
 \includegraphics[width=.75\textwidth]{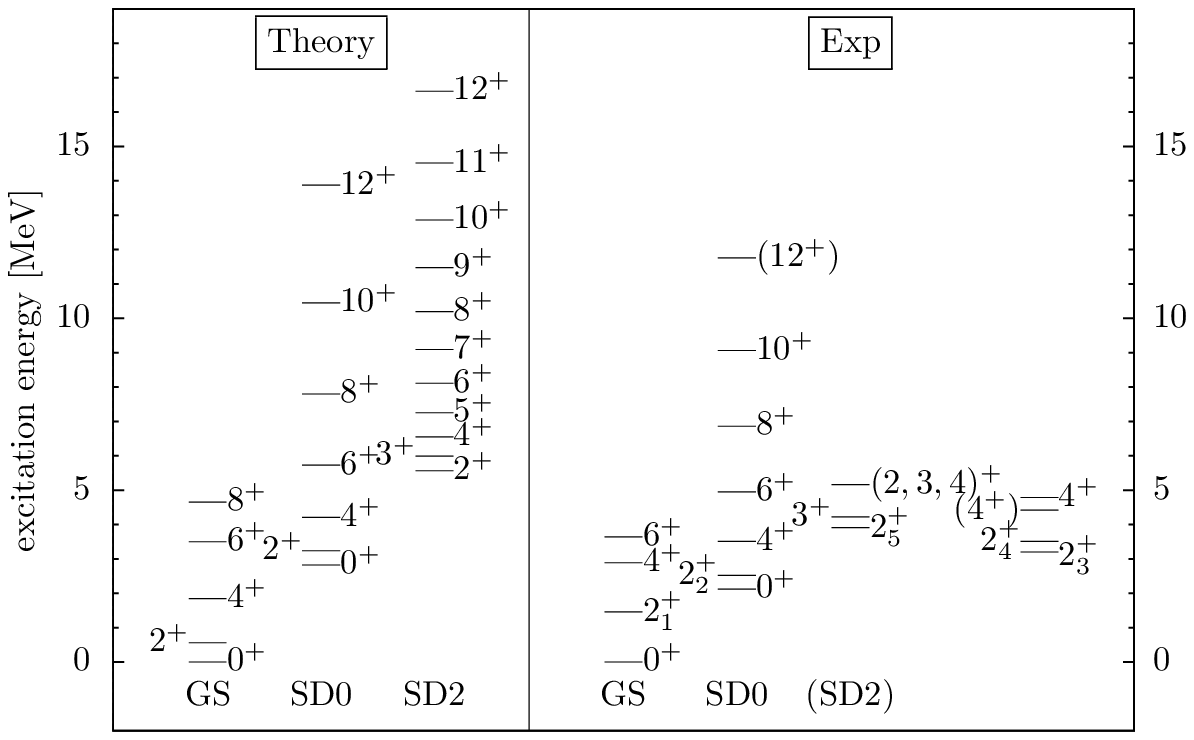}
 \caption{
 Level scheme of $^{40}$Ar is shown.
 In experimental data, the GS, the SD band (SD0), candidate states of the side band of the SD band (SD2), and other low-lying $J^\pi \leq 4^+$ states are listed.
 The experimental data are taken from Refs.~\onlinecite{ideguchi2010} and \onlinecite{Cameron2004293}. 
 }
 \label{fig:level}
\end{figure*}

Figure~\ref{fig:level} shows the experimental and theoretical excitation energies obtained by the GCM.
The $K^\pi = 0^+$ GS band, the $K^\pi = 0^+$ SD band (SD0), and the $K^\pi = 2^+$ SD band (SD2) are obtained by the GCM calculation. The energy levels of the GS and SD0 bands agree with the experimental data. 
The observed $J^\pi = 2^+_5$ (3.92 MeV), $3^+$ (4.23 MeV), and $(2, 3, 4)^+$ (5.17 MeV) states are tentatively assigned as members of the SD2 band in this study because their excitation energies are similar to calculated ones of the SD2 band and the candidates states decay to the members of the SD0 band.
Dominant components of the GS band are the deformed-basis AMD wave functions around the GS minimum on the energy curve, and those of the SD0 and SD2 bands are the basis wave functions around the SD minimum with the triaxial deformations.
Owing to the triaxial deformations around the SD local minimum, the SD2 band appears.
The members of the SD0 and SD2 bands are dominated by the $|K| = 0$ and $2$ components, respectively, projected from a single deformed-basis AMD wave function at $(\beta, \gamma) = (0.478, 11.2^\circ)$ up to high spin states. 
Both the SD0 and SD2 bands show rigid rotor-like behavior. 
The calculated level spacings in the SD0 band members show the rotational band spectra and match experimental data, and those of the SD2 band members are consistent with experimental data. 
The energy gap between the $J^\pi = 0^+_\mathrm{SD0}$ and the $J^\pi = 2^+_\mathrm{SD2}$ states is slightly overestimated, and it may indicate that the triaxiality of the SD states is underestimated. 

The level spacings of the $J^\pi \leq 6^+$ states in the calculated GS band are rotational, but the level spacings between the $J^\pi = 6^+$ and $8^+$ are smaller than those in the case of a rotational band (Fig.~\ref{fig:level}).
The $2 \hbar \omega$ [$(\Delta N_{\rm H.O.}^\pi, \Delta N_{\rm H.O.}^\nu) = (1, 1)$] components mix in the $J^\pi = 8^+$ state; that breaks the rotational spectra.
The calculated GS band deviates from the experimental one and is more rotational.
The experimental GS band may be reproduced by adding sufficient basis wave functions such as $(\Delta N_{\rm H.O.}^\pi, \Delta N_{\rm H.O.}^\nu) = (1, 1)$ components.
The observed $J^\pi = 2^+_3$ (3.21 MeV) and $2_4^+$ (3.51 MeV) states were not obtained perhaps because basis wave functions with a total of $2 \hbar \omega$ configurations are insufficient in the present basis set.

 \begin{table}[tbp]
  \caption{
  Electric quadrupole transition strengths are shown in the Weisskopf unit. 
  $J_i$ and $J_f$ indicate initial and final states, respectively. 
  The blanks under theoretical values indicate that calculated $B(E2)$ values are almost zero. 
  As for the transitions labeled ``---'' under experimental values, $\gamma$ decays have been observed, but multipolarities and strengths have not been measured. 
  The experimental data are taken from Ref.~\onlinecite{Cameron2004293}. 
  The observed $(2, 3, 4)^+$ state at 5.17 MeV is noted as the $4^+_\mathrm{SD2}$.
  }
  \label{tab:BE2}

  \begin{tabular}{cccc}
   \hline
   $J_i$ & $J_f$ & theory & experiments \\
   \hline
   $2^+_\mathrm{GS}$ & $0^+_\mathrm{GS}$ & 6.18 & $9.3 \pm 0.4$ \\
   $4^+_\mathrm{GS}$ & $2^+_\mathrm{GS}$ & 6.65 & $4.8 \pm 1.0$ \\
   $6^+_\mathrm{GS}$ & $4^+_\mathrm{GS}$ & 3.84 & $1.67 \pm 0.05$ \\
   $2^+_\mathrm{SD0}$ & $0^+_\mathrm{SD0}$ & 55.62 &  \\
   $4^+_\mathrm{SD0}$ & $2^+_\mathrm{SD0}$ & 78.54 & $47 \pm 19$ \\
   $6^+_\mathrm{SD0}$ & $4^+_\mathrm{SD0}$ & 84.64 & $70 \pm 30$ \\
   $3^+_\mathrm{SD2}$ & $2^+_\mathrm{SD2}$ & 95.96 &  \\
   $4^+_\mathrm{SD2}$ & $3^+_\mathrm{SD2}$ & 70.75 &  \\
  $4^+_\mathrm{SD2}$ & $2^+_\mathrm{SD2}$ & 31.86 &  \\
  $5^+_\mathrm{SD2}$ & $4^+_\mathrm{SD2}$ & 50.13 &  \\
  $5^+_\mathrm{SD2}$ & $3^+_\mathrm{SD2}$ & 50.72 &  \\
  $6^+_\mathrm{SD2}$ & $5^+_\mathrm{SD2}$ & 31.38 &  \\
  $6^+_\mathrm{SD2}$ & $4^+_\mathrm{SD2}$ & 53.84 &  \\
  \hline
  $2^+_\mathrm{SD2}$ & $2^+_\mathrm{SD0}$ & 1.55 & --- \\
  $2^+_\mathrm{SD2}$ & $0^+_\mathrm{SD0}$ & 0.92 & $1.2 \pm 0.3$\\
  $3^+_\mathrm{SD2}$ & $2^+_\mathrm{SD0}$ & 1.64 & --- \\
  $4^+_\mathrm{SD2}$ & $4^+_\mathrm{SD0}$ & 1.88 & --- \\
  $4^+_\mathrm{SD2}$ & $2^+_\mathrm{SD0}$ & 0.44 &  \\
  \hline
  $0^+_\mathrm{SD0}$ & $2^+_\mathrm{GS}$ &  & $6.2 \pm 2.0$ \\
  $2^+_\mathrm{SD0}$ & $2^+_\mathrm{GS}$ &  & $13 \pm 7$\\
  $2^+_\mathrm{SD0}$ & $0^+_\mathrm{GS}$ &  & $1.32 \pm 0.13$\\
  $4^+_\mathrm{GS}$ & $2^+_\mathrm{SD0}$ &  & $42 \pm 23$ \\
  $4^+_\mathrm{SD0}$ & $2^+_\mathrm{GS}$ &  & $8.1 \pm 1.8$ \\
  $6^+_\mathrm{SD0}$ & $4^+_\mathrm{GS}$ &  & $7 \pm 3$ \\
  \hline
  \\
  \\
  \end{tabular}
 \end{table}

Table~\ref{tab:BE2} shows theoretical and experimental values of electric quadrupole transition strengths $B(E2)$ in the Weisskopf unit. 
The $B(E2)$ values for the intraband transitions of the SD0 and SD2 bands take larger values than those of the GS band, which reflects the large deformations of the SD0 and SD2 bands. 
The $B(E2)$ values for the intraband transitions of the GS and SD0 bands are consistent with experimental data. 
Since the SD0 and SD2 bands are approximately regarded as the $K^\pi = 0^+$ and $2^+$ bands associated with the triaxial superdeformation, the interband
transitions between them have non-negligible values.
The observed $J^\pi = 2^+_5$ (3.92 MeV), $3^+$ (4.23 MeV), and $(2, 3, 4)^+$ (5.17 MeV) decay to the SD0 band members. 
The agreement of the theoretical and experimental $B(E2)$ values of the $2^+_\mathrm{SD2} \rightarrow 0^+_\mathrm{SD0}$ transition supports the band assignment of the SD2 band for the $J^\pi = 2^+_5$ (3.92 MeV) state. 
$B(E2)$ ratios of $B(E2; 2_\mathrm{SD2}^+ \rightarrow 0_\mathrm{SD0}^+) / B(E2; 2_\mathrm{SD0}^+ \rightarrow 0_\mathrm{SD0}^+)$ and $B(E2; 2_\mathrm{SD2}^+ \rightarrow 2_\mathrm{SD0}^+) / B(E2; 2_\mathrm{SD0}^+ \rightarrow 0_\mathrm{SD0}^+)$ are consistent in the case of $\gamma \sim 10^\circ$ with the Davydov-Filippov model \cite{Davydov1958237}, which is a schematic triaxial rotor model. 
To confirm the proposed band assignment of the SD2 band and triaxial deformation of the SD states, more experimental data for intraband transitions of the SD2 band and interband transitions between the SD0 and the SD2 are required. 

To discuss shell effects of protons and neutrons in the SD states of $^{40}$Ar, triaxial aspects of the superdeformation in $^{40}$Ar[$(Z, N) = (18, 22)$] are compared with those in the $N = Z$ isotope ($^{36}$Ar[$(Z, N) = (18, 18)$]) and isotone ($^{44}$Ti[$(Z, N) = (22, 22)$]).
The deformed-basis AMD and HF calculations suggest that the SD states in $^{36}$Ar \cite{Inakura2002261,PTP.121.533} and $^{44}$Ti \cite{Inakura2002261,Kimura200658} form prolate and significantly triaxial shapes, respectively. The theoretically suggested triaxiality $\gamma$ values of the SD states in $^{44}$Ti are approximately $25^\circ$ \cite{Kimura200658} by the deformed-basis AMD, which is consistent with the HF \cite{Inakura2002261}.
The particle-hole configurations of the SD states in $^{36}$Ar and $^{44}$Ti are
considered to be $[(sd)^{-4}(pf)^2]$ and $[(sd)^{-2}(pf)^{4}]$, respectively, for both protons and neutrons, which indicates that the former configuration favors a prolate shape and the latter one induces a triaxial shape.
The SD states in $^{40}$Ar have dominantly the $[(sd)^{-4}(pf)^2]_\pi[(sd)^{-2}(pf)^{4}]_\nu$ configuration, which shows an admixture of shell effects of these configurations.
The triaxiality $\gamma$ values of protons and neutrons of the SD states in $^{40}$Ar are approximately $10^\circ$ (Fig.~\ref{fig:beta-energy}), which is the intermediate value between $\gamma \sim 0^\circ$ and $25^\circ$ for the SD states in $^{36}$Ar and $^{44}$Ti, respectively.
Protons and neutrons may favor similar deformations to increase overlap and gain attractive forces between them. 
Thus, the triaxiality of the SD states in $^{40}$Ar is understood by a competition of a prolate shape favored by proton shell effects for $Z = 18$ and a significantly triaxial shape induced by neutron shell effects for $N = 22$.

In conclusion, the structure of the SD states in $^{40}$Ar has been investigated with the deformed-basis AMD and the GCM.
The SD states in $^{40}$Ar form triaxial shapes. 
Owing to the triaxial deformation, the $K^\pi = 2^+$ SD band exists. 
Calculated $B(E2)$ values for intraband transition of the GS and SD0 bands are consistent with experimental data. 
The triaxiality $\gamma$ of the SD states is understood by the competition of deformations favored by their proton and neutron structures. 
To understand details of nuclear structures, it is necessary to take into account triaxiality. 

Numerical calculations were conducted on the RIKEN Cluster of Clusters (RICC) and the supercomputer SX9 in the Research Center for Nuclear Physics, Osaka University. 
The authors thank Dr.~N.~Itagaki in the Department of Physics, Faculty of Science, The University of Tokyo and Dr.~T.~Yoshida in the Center for Nuclear Study, The University of Tokyo, for fruitful discussions. 

\bibliography{40Ar_vs2}

\end{document}